\documentclass[runningheads]{llncs}
\usepackage{graphicx}
\usepackage{amsmath}
\usepackage{caption}
\newcommand{\tabincell}[2]{\begin{tabular}{@{}#1@{}}#2\end{tabular}} %%table
\usepackage{amssymb} %mathbb
\usepackage{booktabs}
\usepackage[colorlinks,
            linkcolor=blue,       %%
            anchorcolor=blue,  %%
            citecolor=black,        %%
            ]{hyperref}
\begin{document}
\title{Chest X-ray Image Classification:\\ A Causal Perspective}
% %%%%%%%%%%%%%%%%%%%%%%%%%%%%%%%%%%%%%%%%%%%%%%%%%%%%%%%%%%%%%%%%%%%%%%%
% %\titlerunning{Abbreviated paper title}
% % If the paper title is too long for the running head, you can set
% % an abbreviated paper title here
% %\and Third Author\inst{3}\orcidID{2222--3333-4444-5555}
\author{Weizhi Nie\inst{1} \and
Chen Zhang\inst{1} \and
Dan Song*\inst{1} \and
Lina Zhao\inst{2} \and
Yunpeng Bai\inst{3} \and
Keliang Xie\inst{4} \and
Anan Liu\inst{1} }
\authorrunning{W. Nie et al.}
% % First names are abbreviated in the running head.
% % If there are more than two authors, 'et al.' is used.
% %
\institute{Tianjin University, Tianjin 300072, China\\
\email{\{weizhinie,zhangchen001,dan.song\}@tju.edu.cn, anan0422@gmail.com}\\
\and
Department of Critical Care Medicine, Tianjin Medical University General Hospital, Tianjin 300052, China\\
\email{18240198229@163.com}
\and
Department of Cardiac Surgery, Chest Hospital, Tianjin University, and Clinical school of Thoracic, Tianjin Medical University, Tianjin 300052, China
\email{oliverwhite@126.com}\\
% \url{http://www.springer.com/gp/computer-science/lncs} \and
\and
Department of Critical Care Medicine, Department of Anesthesiology, and Tianjin Institute of Anesthesiology, Tianjin Medical University General Hospital, Tianjin 300052, China\\
\email{xiekeliang2009@hotmail.com}
}
% %
% \maketitle              % typeset the header of the contribution
% %%%%%%%%%%%%%%%%%%%%%%%%%%%%%%%%%%%%%%%%%%%%%%%%%%%%%%%%%%%%%%%%%%%%%%%%%

%%%%%%%%%%%%%%%%%%%%%%%%%%%%%%%%%%%%%%%%%%%%%%%%%%%%%%%%%%%%%%%%%%%%%%%
% \author{Anonymous}

% \authorrunning{}

% \institute{}
% \email{}
% \and
% Springer Heidelberg, Tiergartenstr. 17, 69121 Heidelberg, Germany
% \email{lncs@springer.com}\\
% % \url{http://www.springer.com/gp/computer-science/lncs} \and
% \and
% ABC Institute, Rupert-Karls-University Heidelberg, Heidelberg, Germany\\
% \email{\{abc,lncs\}@uni-heidelberg.de}}
%
\maketitle              % typeset the header of the contribution
%%%%%%%%%%%%%%%%%%%%%%%%%%%%%%%%%%%%%%%%%%%%%%%%%%%%%%%%%%%%%%%%%%%%%%%%%

\begin{abstract}
The chest X-ray (CXR) is one of the most common and easy-to-get medical tests used to diagnose common diseases of the chest. Recently, many deep learning-based methods have been proposed that are capable of effectively classifying CXRs. Even though these techniques have worked quite well, it is difficult to establish whether what these algorithms actually learn is the cause-and-effect link between diseases and their causes or just how to map labels to photos.
In this paper, we propose a causal approach to address the CXR classification problem, which constructs a structural causal model (SCM) and uses the backdoor adjustment to select effective visual information for CXR classification. Specially, we design different probability optimization functions to eliminate the influence of confounders on the learning of real causality. Experimental results demonstrate that our proposed method outperforms the open-source NIH ChestX-ray14 in terms of classification performance.
\end{abstract}

\section{Introduction}
As a non-invasive test, the chest X-ray (CXR) is often used by doctors to diagnose diseases of the thorax. In clinical practice, the acquisition of diagnostic results of CXR is always interpreted by professional radiologists, which is expensive in terms of time and easily affected by the individual's medical abilities~\cite{brady2012discrepancy}. Thus, some researchers tend to find some automated and accurate CXR classification technology based on machine learning, which can help doctors to make better diagnoses
\cite{wang2017chestx,irvin2019chexpert,rocha2022attention,ke2021chextransfer,saleem2021classification}. However, there are some inherent problems with CXR images that are difficult to solve, such as high interclass similarity~\cite{rajaraman2020training}, dirty atypical data, complex symbiotic relationships between diseases~\cite{wang2017chestx}, and long-tailed or imbalanced data distribution~\cite{zhang2021deep}. 
%For example, we know it is a challenging task to classify some rare categories of CXR images, but if these images have some kind of letter marker or medical device in them just because the same disease may be treated similarly, the deep model tends to tackle the task by aforementioned confounding features. 

\begin{figure}[h]
\centering\includegraphics[width=0.7\textwidth]{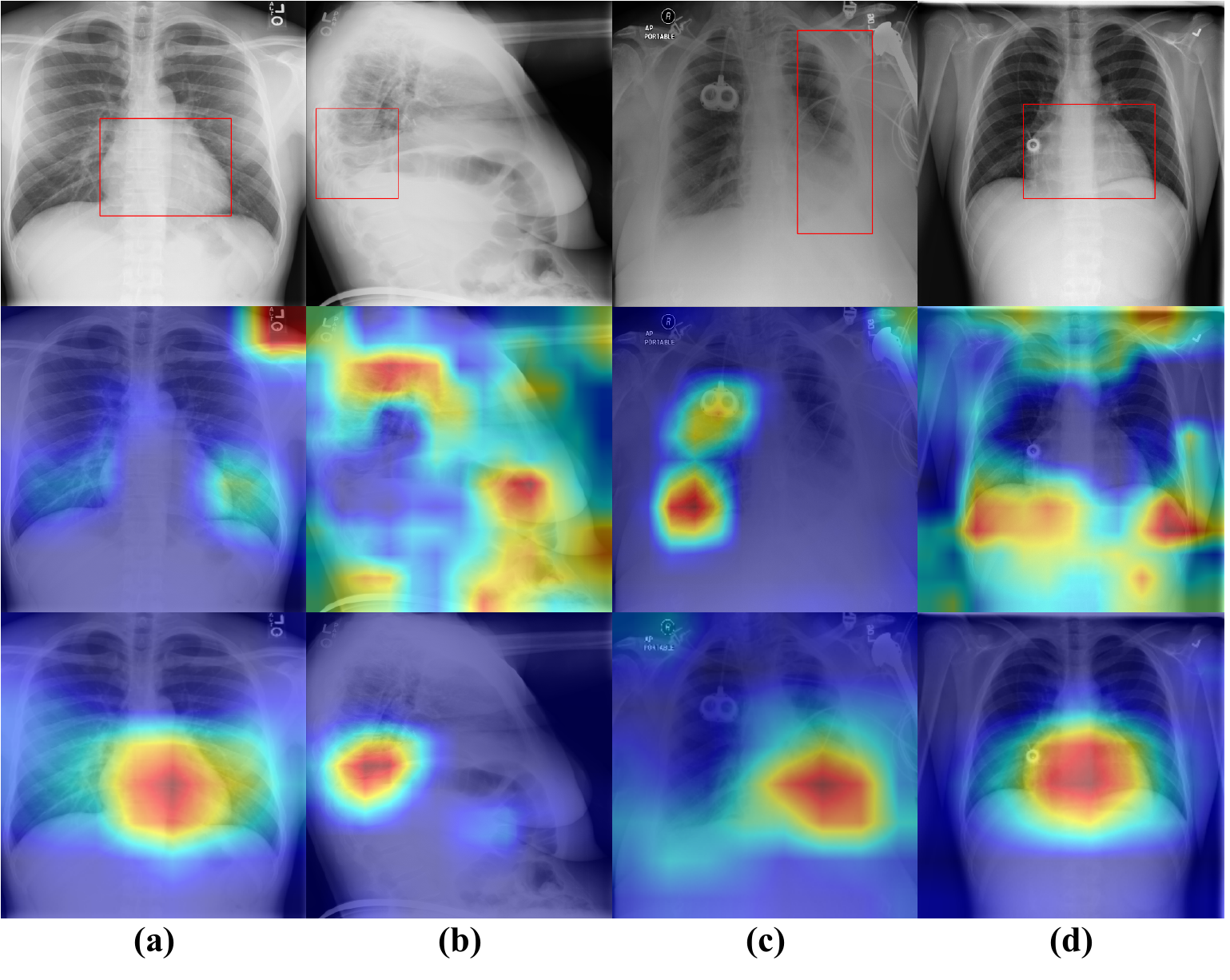}
\caption{Some tough cases in the data set. Each column is the same CXR image, and each row from top to bottom shows the original image with a pathological bounding box, weighted heat maps of traditional CNN-based deep learning, and our proposed method, in that order. Four difficult situations such as (a): letters on images, (b): irregular images, (c): medical devices on images, and (d): easily confused between classes.} \label{example}
\end{figure}

Some examples are shown in Fig.~\ref{example} from the NIH dataset, we can find previous methods performed not stable when dealing with some tough cases. For example, the label of Fig.~\ref{example}(d) is cardiomegaly but the predicting results generated by a CNN-based model are infiltration, which fits the statistical pattern of symbiosis between these two pathologies~\cite {wang2017chestx}. 
%Besides, traditional deep learning-based models inevitably suffer from a performance drop in an out-of-distribution (OOD) scenario simply because the probability distributions of the training and test data are different. 
%In addition, t
To the black-box nature of deep learning, even if their proposed model has a decent performance, it is also difficult to determine whether what is learned is true causality.
%the lack of interpretability still can not win the trust of specialists. 
Unfortunately, some recent efforts such as \cite{rocha2022attention,liu2019sdfn} already notice part of the above problems but only try to solve it by data pre-processing or designing complicated model, they fail to let the deep model capture real causality.

In order to effectively solve the above problems, we model the CXR image classification task from a causal perspective. We sort out the relationships among the causal feature, the confounding feature, and the classification result. In a nutshell, our basic idea is ``borrow from others.'' We also stick with letters in an image as an example. If part of the image is marked with letters, this situation affects the classification of the unmarked image, because we think of the letters as a confounding element, When we borrow the mark from the unmarked image, it is equivalent to eliminating the confounding effect. The same is true for other confounding assumptions we mentioned.

Towards this end, we utilize causal inference to minimize the confounding effect and maximize the causal effect while achieving stable and decent performance. Specifically, we first utilize traditional CNN-based modules to extract the feature from the input CXR images, and then apply Transformer decoder~\cite{vaswani2017attention} based cross-attention mechanism to produce the estimations of the causal and latent confounding features from the feature maps. After that, we can parameterize the backdoor adjustment in the causal theory\cite{pearl2000models}, which combines every causal estimation with different confounding estimations and encourages these combinations to remain a stable classification performance via the idea of ``borrow from others''. It tends to facilitate the invariance between the causal patterns and the classification results. 
%in spite of the variation in the ``useless for classification'' parts or OOD problems.

We apply the method to different data sets and the experimental results demonstrate the performance and superiority of our approach. Our contributions can be summarized as follows:
\begin{itemize}
    \item We take a casual look at the chest X-ray images' multi-label classification problem and model the disordered or easily-confused part of an image as the confounder.
    \item We propose a framework based on the guideline of backdoor adjustment and presented a novel strategy for chest X-ray image classification. It allows a properly designed model to exploit real and stable causal features while removing the effects of filtrable confounding patterns.
    \item Extensive experiments on two large-scale public datasets justify the effectiveness of our proposed method. More visualizations with detailed analysis demonstrate the interpretability and rationalization of our proposed method.
\end{itemize}

\section{Methodology}
In this section, we first define the causal model, then identify the strategies to eliminate confounding effects.

\subsection{A Causal View on CXR Images}
\begin{figure}
\centering\includegraphics[width=0.7\textwidth]{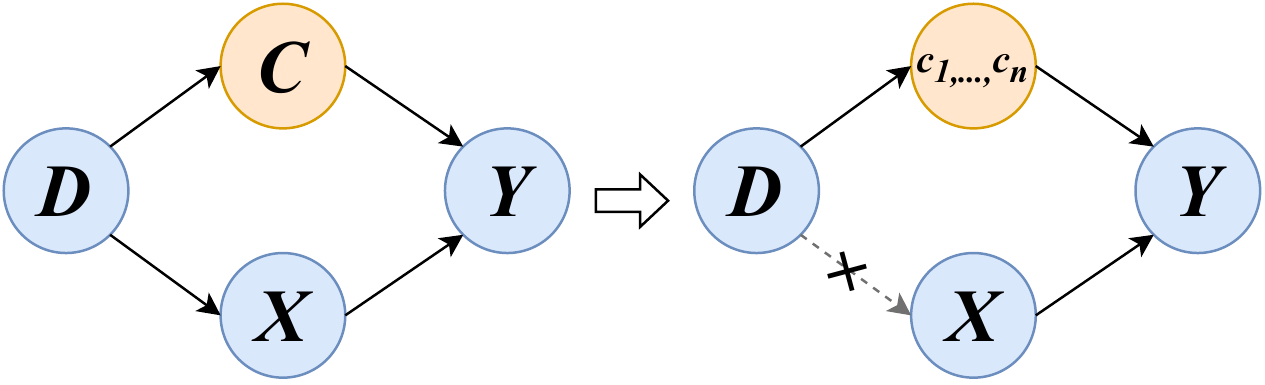}
\caption{Structural causal model for CXR image classification. ``D'' is the input data, ``C'' denotes the confounding features, ``X'' is the causal features and ``Y'' is the prediction results.} \label{fig2}
\end{figure}

From the above discussion, we construct a Structural Causal Model (SCM)~\cite{glymour2016causal}in Fig.~\ref{fig2}(a) to solve the spurious correlation problems in CXR. It contains the causalities about four elements: Input CXR image data $D$, confounding feature $C$, causal feature $X$, and prediction $Y$, where the arrows between elements stand for cause and effect: cause $\rightarrow$ effect. We have the following explanations for the SCM in our task:
\begin{itemize}
    \item[$\bullet$] $C \leftarrow D \rightarrow S$: $X$ denotes the causal feature which really contributes to the diagnosis, whereas $C$ denotes the confounding feature which may mislead the diagnosis and usually caused by data bias and other complex situations mentioned above. The two arrows can be seen as the feature extraction process. Apparently, $C$ and $X$ usually coexist in the medical image data $D$, these causal effects are built naturally.
    \item[$\bullet$] $C \rightarrow Y \leftarrow X$: We denote $Y$ as the classification result which should have been caused only by $X$ but inevitably disturbed by confounding features. The two arrows can be implemented by classifiers. 
\end{itemize}

The goal of the classification model should capture the true causality between the causal feature $X$ and the diagnostic result $Y$, avoiding the influence of the confounding feature $C$. For example, we hope in some complex cases, the model will diagnose via real pathology features rather than letters or medical devices in the input CXR image. However, the conventional correlation $P(Y|X)$ fails to achieve that because of the backdoor path~\cite{pearl2014interpretation} $X \leftarrow D \rightarrow C \rightarrow Y$ between $X$ and $Y$. Therefore, we choose to apply the causal intervention to cut off the backdoor path and use $P(Y|do(X))$ to replace $P(Y|X)$, so the model has the ability to exploit causal features.

\subsection{Causal Intervention via Backdoor Adjustment}
In this section, we propose to use the backdoor adjustment~\cite{glymour2016causal} to implement $P(Y|do(X))$ and eliminate the backdoor path, which is shown in Fig.~\ref{fig2}(b). The backdoor adjustment assumes that we can observe and stratify the confounders, \textit{i.e.}, $C=\{c_1,c_2,...,c_n\}$, where each $c$ is a stratification of the confounder feature. We can then exploit the powerful \textbf{do-calculus} on causal feature $X$ by estimating $P_{b}(Y|X) = P(Y|do(X))$, where the subscript $b$ denotes the backdoor adjustment on the SCM.

Causal theory~\cite{pearl2000models} provides us with three key conclusions: 
\begin{itemize}
    \item[$\bullet$] $P(c) = P_{b}(c)$: the marginal probability is invariant under the intervention, because $C$ will remain unchanged when cutting the link between $D$ and $X$ as shown in Fig.~\ref{fig2}(b).
    \item[$\bullet$] $P_{b}(Y|X,c) = P(Y|X,c)$: $Y$'s response to $X$ and $C$ has no connection with the causal effect between $X$ and $C$.
    \item[$\bullet$] $P_{b}(c|X) = P_{b}(c)$: $X$ and $C$ are independent after backdoor adjustment.
\end{itemize} 

Based on the conclusions, the backdoor adjustment for the SCM in Fig.~\ref{fig2}(a) is:
\begin{equation}
\begin{aligned}
P(Y|do(X))
&=P_{b}(Y|X)=\sum_{c \in \mathcal{C}}P_{b}(Y|X,c)P_{b}(c|X)\\
&=\sum_{c \in \mathcal{C}}P_{b}(Y|X,c)P_{b}(c)=\sum_{c \in \mathcal{C}}P(Y|X,c)P(c),
\end{aligned}
\label{bd}
\end{equation}
where $\mathcal{C}$ denotes the confounder set, $P(c)$ is the prior probability of $c$. Then, we approximate the formula by a random sample operation which will be detailed next. 

\subsection{Training Object}
\begin{figure}
\includegraphics[width=\textwidth]{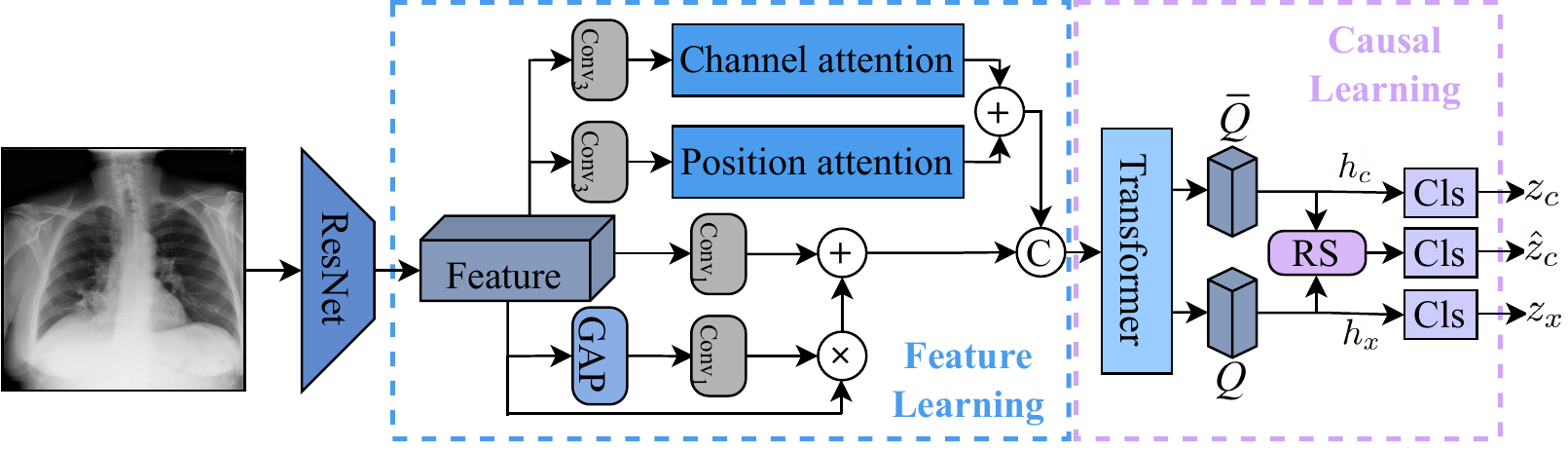}
\caption{Overview of our network. Firstly, we apply CNN with modified attention to extract the image feature, where the $n$ in Conv$_n$ denotes the kernel size of the convolutional operation, ``+'', ``$\times$'', and ``C'' denote add, multiply, and concatenate operations, respectively. ``GAP'' means global average pooling, ``RS'' is the random sample operation, and ``Cls'' denotes the classifier. The cross-attention module inside the transformer decoder disentangles the causal and confounding feature, then we can apply parameterized backdoor adjustment to achieve causal inference.} \label{fw}
\end{figure}

Till now, we need to provide the implementations of Eq.(\ref{bd}) in a parameterized method to fit the deep learning model. However, in the medical scenario, $\mathcal{C}$ is complicated and hard to obtain, so we simplify the problem and assume a uniform distribution of confounders. 
Traditionally, when we want to drive the deep model to learn useful knowledge, we are always extremely dependent on the properly designed loss function. 
Then, towards effective backdoor adjustment, we utilize different loss functions to drive our deep model to learn causal and spurious features respectively. Fig.~\ref{fw} illustrates the proposed network. Note that the channel and position attention is implemented by adopting an efficient variant of self-attention \cite{pan2022integration}. We will break the whole framework down in detail below.

Given an image $x \in \mathbb{R}^{H_0 \times W_0 \times 3}$ as input, we extract its spatial feature $F \in \mathbb{R}^{H \times W \times d}$ using the backbone, where $H_0 \times W_0$, $H \times W$ represent the height and width of the CXR image and the feature map respectively, and $d$ denotes the hidden dimension of the network. Then, we adopt zero-initialized $Q_0 \in \mathbb{R}^{C \times d}$ as the queries in the cross-attention module inside the transformer, each decoder layer $l$ updates the queries $Q_{l-1}$ from its previous layer. Here, we denote $Q$ as the causal feature and $\overline{Q}$ as the confounding feature as follows:
\begin{equation}
\begin{aligned}
&Q_l= softmax(\widetilde{Q}_{l-1}\widetilde{F}/\sqrt{dim_{\widetilde{F}}})F,\\
&\overline{Q}_l= (1 - softmax(\widetilde{Q}_{l-1}\widetilde{F}/\sqrt{dim_F}))F,
% &F_l=FFN(LN(\dot{F}_l))+\dot{F}_l,
\end{aligned}
\label{decoder}
\end{equation}
where the tilde in $\widetilde{F}$ means the feature with position encodings, the disentangled features yield two branches, which can be fed separately into a point-wise Multi-layer perceptron (MLP) network and get corresponding classification logits via a sigmoid function.

\subsubsection{Disentanglement.} As shown in Fig.~\ref{fw}, we try to impel the model to learn both causal and confounding features via the designed model structure and loss function. 
Specifically, we adopt a CNN-based model to extract the feature of input images, then capture the causal feature and confounding feature by cross-attention mechanism. Thus we can make the prediction via MLP and classifiers:
\begin{equation}
\begin{aligned}
&h_c= MLP_{confounding}(\overline{Q_l}), z_c= \Phi_c(h_c),\\
&h_x= MLP_{causal}(Q_l), z_x= \Phi_x(h_x),
\end{aligned}
\label{feature}
\end{equation}
where $h \in \mathbb{R}^{d \times C}$, $C$ is the number of categories,  $\Phi(\cdot)$ represents classifier, $z$ denotes logits.

The causal part aims to estimate the really useful feature, so we apply the supervised classification loss in a cross-entropy format:
\begin{equation}
% \sum^{C}_{c=1}
\mathcal{L}_{sl}= -\frac{1}{\lvert D \rvert}\sum_{d \in D} y^\top\log(z_x),
\label{loss1}
\end{equation}
where $d$ is a sample and $D$ is the training data, $y$ is the corresponding label. The confounding part is unwanted for classification, so we follow the work in CAL~\cite{sui2022causal} and push its prediction equally to all categories, then the confounding loss is defined as:
\begin{equation}
\mathcal{L}_{conf}= -\frac{1}{\lvert D \rvert}\sum_{d \in D} KL(y_{uniform}, z_c),
\label{loss2}
\end{equation}
where KL is the KL-Divergence, $y_{uniform}$ denotes a uniform distribution. We optimize the above two losses and can effectively disentangle causal and confounding features.

\subsubsection{Causal intervention.} The idea of the backdoor adjustment formula in Eq.(\ref{bd}) is to stratify the confounder and combine confounding and causal features manually, which is also the implementation of the random sample in Fig.~\ref{fw}. For this propose, we stratify the extracted confounding feature and random add it to the other CXR images' feature to be classified shown in Eq.(\ref{hatzc}), and get a ``intervened graph'', then we have the following loss guided by causal inference:
\begin{equation}
\hat{z}_c= \Phi(h_x+\hat{h}_c),
\label{hatzc}
\end{equation}
\begin{equation}
\mathcal{L}_{bd}= -\frac{1}{\lvert D \rvert \cdot \lvert \hat{D} \rvert}\sum_{d \in D} \sum_{\hat{d} \in \hat{D}} y^\top\log(\hat{z}_c),
\label{loss3}
\end{equation}
where $\hat{z}_c$ is the prediction from a classifier on the ``intervened graph'', $\hat{h}_c$ is the stratification feature via Eq.(\ref{feature}), $\hat{D}$ is the estimated stratification set contains trivial features. The objective of our framework can be defined as the sum of the losses:
\begin{equation}
\mathcal{L}= \mathcal{L}_{sl}+\alpha_1\mathcal{L}_{conf}+\alpha_2\mathcal{L}_{bd},
\label{loss}
\end{equation}
where $\alpha_1$ and $\alpha_2$ are hyper-parameters, which decide how powerful disentanglement and backdoor adjustment are. It pushes the prediction stable because of the shared image features according to our detailed experimental results in the next section.

\section{Experiments}
\subsection{Experimental Setup}
We evaluate the common thoracic diseases classification performance on the NIH ChestX-ray14~\cite{wang2017chestx}
% and CheXpert~\cite{irvin2019chexpert} 
data set, which consists of 112,120 frontal-view CXR images with 14 diseases and we follow the official data split for a fair comparison. 
% The latter consists of 224,316 frontal and lateral view CXR images of 14 diseases which are more challenging, we select the lower resolution of 320 $\times$ 320 and also follow the official split of data.

In our experiments, we adopt ResNet101~\cite{he2016deep} as the backbone. Our experiment is operated by using NVIDIA GeForce RTX 3090 with 24GB memory. We use the Adam~\cite{kingma2014adam} optimizer with a weight decay of 1$e$-2 and the max learning rate is 1$e$-3. On the NIH data set, we resize the original images to 512 $\times$ 512 as the input.
% and 320 $\times$ 320 on the CheXpert data set for a fair comparison. 
We evaluate the classification performance of our method with the area under the ROC curve (AUC) for the whole test set.

% \subsection{NIH Chest-Xray14 Dataset}
\subsection{Results and Analysis}
Table.~\ref{tb1} illustrates the overall performance of the NIH Chest-Xray14 dataset of our proposed method compared with other previous state-of-art works, the best performance of each pathology is shown in bold. From the experiments on the NIH data set, we can conclude that we eliminate some spurious relationships inner and between CXR images from the classification results. Specifically, we can find that we are not only making progress in most categories but also dealing with some pathologies with high symbiotic dependence~\cite{wang2017chestx} such as cardiomegaly and infiltration.
% Note that ``Cardiomegaly'', ``Emphysema'', ``Edema'', ``Fibrosis'', ``Pneumonia'', and ``Hernia'' only account for 7.68\% of the total pathologies, we are also delighted to discover that the long-tail problem has been improved via our proposed causal inference-based method. 
The visualization results in Fig.~\ref{example} prove that the issues raised were addressed. 

\begin{table*}[t]
\caption{Comparation of AUC scores with previous SOTA works. We report the AUC with a 95\% confidence interval (CI) of our method. }
  \centering
    \begin{tabular}{c|c|c|c|c|c}
    \toprule
    % Abnormality   & DNetLoc\cite{guendel2018learning} & Xi \textit{et al.}\cite{ouyang2020learning} & ImageGCN\cite{mao2022imagegcn} &DGFN\cite{gong2021deformable} & Ours \\
    \tabincell{c}{Abnormality}&\tabincell{c}{DNetLoc\\\cite{guendel2018learning}  }&\tabincell{c}{Xi \textit{et al.}\\\cite{ouyang2020learning}}&\tabincell{c}{ImageGCN\\\cite{mao2022imagegcn}}&\tabincell{c}{DGFN\\\cite{gong2021deformable}}&\tabincell{c}{Ours}\\
    \midrule
    % \midrule
    Atelectasis   & 0.77  & 0.77  & 0.80 & \textbf{0.82} & 0.81 (0.81, 0.82) \\
    % \midrule
    Cardiomegaly     & 0.88  & 0.87  & 0.89 &0.93& \textbf{0.94} (0.93, 0.95) \\
    % \midrule
    Effusion    & 0.83   & 0.83  & 0.87 &0.88 & \textbf{0.91} (0.91, 0.92) \\
    % \midrule
    Infiltration   & 0.71   & 0.71  & 0.70 & \textbf{0.75} & \textbf{0.75} (0.74, 0.77) \\
    % \midrule
    Mass    & 0.82   & 0.83  & 0.84 & 0.88 & \textbf{0.89} (0.88, 0.90)\\
    % \midrule
    Nodule    & 0.76    & \textbf{0.79}  & 0.77 & \textbf{0.79} & 0.76 (0.74, 0.79) \\
    % \midrule
    Pneumonia    & 0.73    & \textbf{0.82}  & 0.72 & 0.78& \textbf{0.82} (0.80, 0.83) \\
    % \midrule
    Pneumothorax    & 0.85   & 0.88  &0.90  &0.89 & \textbf{0.91} (0.91, 0.93)\\
    % \midrule
    Consolidation     & 0.75  & 0.74  & 0.80 & 0.81 & \textbf{0.82} (0.81, 0.83)\\
    % \midrule
    Edema    & 0.84   & 0.84  & 0.88 &0.89 & \textbf{0.90} (0.89, 0.90)\\
    % \midrule
    Emphysema    & 0.90   & \textbf{0.94}  & 0.92 & \textbf{0.94} & \textbf{0.94} (0.93, 0.95) \\
    % \midrule
    Fibrosis   & 0.82    & 0.83  & 0.83 & 0.82 & \textbf{0.84} (0.84, 0.85) \\
    % \midrule
    Pleural\_Thicken     & 0.76  & 0.79  & 0.79 & \textbf{0.81} & 0.77 (0.75, 0.78)\\
    % \midrule
    Hernia   & 0.90 & 0.91  & \textbf{0.94} & 0.92 & \textbf{0.94} (0.92, 0.95) \\
    \midrule
    % \midrule
    Mean AUC     & 0.807   & 0.819  & 0.832 &0.850 & \textbf{0.857} (0.849, 0.864)\\
    \bottomrule
    \end{tabular}%
  \label{tb1}%
\end{table*}%

\begin{figure}[!ht]
    \centering
	\begin{minipage}{0.4\linewidth}
		\centering
		\includegraphics[width=\linewidth]{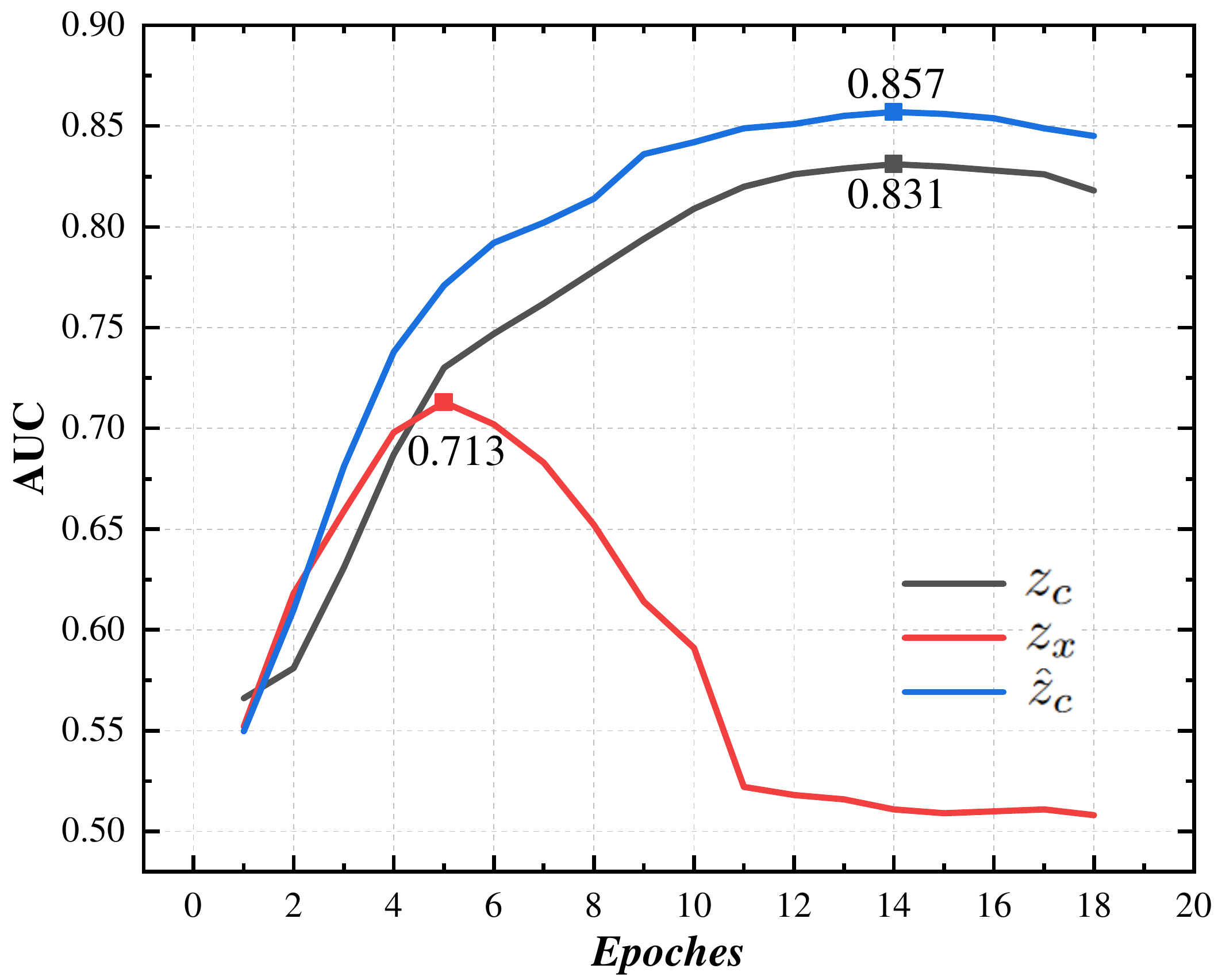}
        \caption{Fluctuation of classification effect of three classifiers.}
		\label{classifier}
	\end{minipage}
	%\qquad
	% \hfill
	\begin{minipage}{0.4\linewidth}
		\centering
        \captionof{table}{Ablation study on NIH data set.}
        \begin{tabular}{c|c|c|c}
        \toprule    Model &\tabincell{c}{Feature\\Learning }&\tabincell{c}{Causal\\Learning }& AUC    \\
        \midrule   
        1     & -     & -     & 0.812       \\
        2     & -     & +     & 0.833      \\
        3     & +     & -     & 0.824       \\
        4     & +     & +     & \textbf{0.857}       \\
        \bottomrule    \end{tabular}%
        \label{ablation}%
    \end{minipage}
\end{figure}

Ablation studies on the NIH data set are shown in Table.~\ref{ablation}. Where ``+'' denotes utilize the module whereas ``-'' denotes remove the module. We demonstrate the efficiency of our method from the ablation study, we can find that our feature extraction and causal learning module play significant roles, respectively.
% For ``Feature learning'' and ``Causal learning'' modules, the AUC for ``--'', ``-+'', ``+-'', and ``++'' are 0.812, 0.833, 0.824, and 0.857, respectively.
Besides, during the training process, Fig.~\ref{classifier} shows the fluctuation of the classification effect of three classifiers, where the three lines in the diagram correspond to the three classifiers in Fig.~\ref{fw}. We can find the performance of the confounding classifier goes up at first and then down. At the same time, the other two classifiers' performance increased gradually, which is in line with our expectations. Our proposed causal learning framework successfully discards the adverse effect of confounding features and makes the prediction stable.

\section{Conclusion}
In conclusion, we present a novel causal inference-based chest X-ray image multi-label classification framework from a causal perspective, which comprising a feature learning module and a backdoor adjustment-based causal inference module. We find that previous deep learning based strategies are prone to make the final prediction via some spurious correlation, which plays a confounder role then damage the performance of the model.  We evaluate our proposed method on the public data set, experimental results indicate that our proposed framework and method are superior to previous state-of-the-art methods.

% This work was supported in part by the National Natural Science Foundation of China (61872267) and the Natural Science Foundation of Tianjin (16JCZDJC31100, 16JCZDJC31100).

%
% ---- Bibliography ----
%
% BibTeX users should specify bibliography style 'splncs04'.
% References will then be sorted and formatted in the correct style.
%
\bibliographystyle{splncs04}
\bibliography{mybibliography}
%
% \begin{thebibliography}{8}
% \bibitem{ref_article1}
% Author, F.: Article title. Journal \textbf{2}(5), 99--110 (2016)

% \bibitem{ref_lncs1}
% Author, F., Author, S.: Title of a proceedings paper. In: Editor,
% F., Editor, S. (eds.) CONFERENCE 2016, LNCS, vol. 9999, pp. 1--13.
% Springer, Heidelberg (2016). \doi{10.10007/1234567890}                                                                                                                                                              

% \bibitem{ref_book1}
% Author, F., Author, S., Author, T.: Book title. 2nd edn. Publisher,
% Location (1999)

% \bibitem{ref_proc1}
% Author, A.-B.: Contribution title. In: 9th International Proceedings
% on Proceedings, pp. 1--2. Publisher, Location (2010)

% \bibitem{ref_url1}
% LNCS Homepage, \url{http://www.springer.com/lncs}. Last accessed 4
% Oct 2017
% \end{thebibliography}
\end{document}